\documentclass[twocolumn,showpacs,aip,jap,preprintnumbers,
amssymb,amsmath]{revtex4}
\usepackage{amssymb,amsmath}
\usepackage{graphicx,float}
\usepackage{bm}
\usepackage{array}
\begin{document}
\title{Compressible spherical dipolar glass model 
of relaxor ferroelectrics}

\author{R. Pirc}
\email{rasa.pirc@ijs.si}
\author{Z. Kutnjak}
\author{N. Novak}
\affiliation{Jo\v zef Stefan Institute,  P.O. Box 3000, 1001 Ljubljana, 
Slovenia}

\date{\today}

\begin{abstract}
The interactions between the dielectric polarization 
and the fluctuations of the strain (stress) tensor in relaxor 
ferroelectrics are shown to give rise to the anisotropy of the 
anharmonic $P^4$-term in the Landau-type free energy, however, 
the harmonic $P^2$-term is still properly described 
by the rigid spherical random bond--random field model. 
These are the essential features of the compressible spherical 
dipolar glass model, which is used to calculate the singularities 
of the specific heat near field-induced critical points. The
results agree with recent high-resolution calorimetric 
experiments in PMN [110]. 
\end{abstract}


\maketitle
~~~~


\section{Introduction}

Relaxor ferroelectrics (relaxors) exhibit a variety of physical properties 
which are interesting for numerous practical applications, such as 
tunable capacitors, ultrasonic transducers, actuators, and pyroelectric 
detectors \cite{KU}. Sometimes relaxors are regarded as a 
subgroup of incipient ferroelectrics in view of the fact that they 
do not possess a polarized long-range ordered phase in zero applied 
electric field. However, in contrast to normal incipient ferroelectrics, 
relaxors undergo a freezing transition into a nonergodic glass-like 
phase below the so-called freezing temperature. If the relaxor is
slowly cooled in a nonzero electric field $E$, it will pass through
a sequence of quasi-stationary states. Thus, in order to stay close
to thermal equilibrium, the experimental scale should increase 
steadily as the temperature is lowered. The corresponding $E$-$T$ phase 
diagram is shown in Fig.~1 for the case of PbMg$_{1/3}$Nb$_{2/3}$O$_3$ 
(PMN) in a field along the [110] direction \cite{NWPK}. The solid line
in Fig.~1 separates the field-cooled dipolar glass phase from the
field-induced long-range correlated ferroelectric phase. Similarly,
the dotted line represents the boundary between the ergodic paraelectric 
phase and the frozen-in nonergodic dipolar glass phase. On approaching 
this line from the right, the longest dielectric 
relaxation time $\tau(E,T)$ diverges according to the Vogel-Fulcher 
law \cite{LKFP,B1}, reflecting the random character of the relaxor state.

Dielectric experiments in PMN [111] \cite{KLP,K2} 
have shown that there is no frequency dispersion of the dielectric
susceptibility in the region above the solid line, indicating that 
the relaxation times are finite and the system is ergodic. The 
transitions across the solid line, indicated by the arrows, are 
all first order and are characterized by a jump in the 
polarization $P=P(E,T)$. As one moves towards higher temperatures, 
the size of the jump becomes smaller and finally disappears at a 
liquid-vapor-type critical point $T_{CP},E_{CP}$, where the transition 
is second order \cite{K5}. Beyond this point, the relaxor is 
in a supercritical state characterized by a smooth evolution of 
$P(E,T)$ and of the field-dependent dielectric susceptibility 
$\chi(E,T)=(\partial P/\partial E)_T$. 
The dashed line marks the positions of the maxima of $\chi(E,T)$ 
(Widom line).


\begin{figure}
\resizebox{20.4pc}{!}{\includegraphics{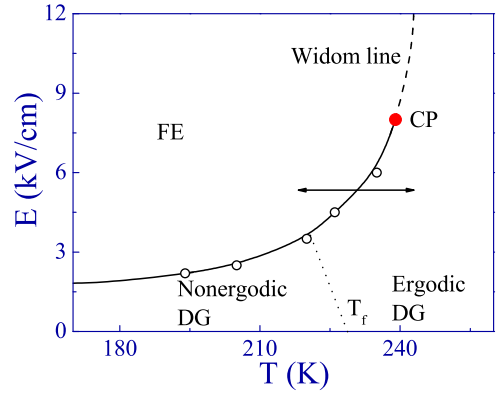}}
\caption{Phase diagram for a relaxor with $b<0$. The solid line separates
the dipolar glass (DG) phase from the field-induced ferroelectric (FE) 
phase. The arrows indicate the direction of the field-cooling (-heating) 
process. Dashed line: Supercritical regime. Dotted line: 
Freezing line separating the ergodic dipolar glass phase from the 
nonergodic one. Open circles: Data from Ref.~\cite{NWPK}.}
\label{fig1}
\end{figure} 


\section{Polarization-stress coupling in Landau free energy}

It had been suggested earlier \cite{PKBZ} that when dealing with 
quasi-equilibrium states as in Fig.~1, the relaxor can be described 
in terms of a Landau theory based on the free energy density 
\begin{equation}
F=F_0+\frac{1}{2}\,a\,P^2
+\frac{1}{4}\,b\,P^4+\frac{1}{6}\,c\,P^6+\cdots -EP.
\label{f}
\end{equation}
For simplicity we are dealing with a scalar order parameter
$P=P(E,T)$, corresponding to the polarization vector along one of the
symmetry directions in the crystal, i.e., [100], [110], or [111]
in a system with average cubic symmetry. For an oblique direction
of the electric field $\vec{E}$ the quartic term should be written as
\begin{equation}
F_4=\frac{1}{4}\,b_{ijkl}\,P_iP_jP_kP_l,
\label{f4}
\end{equation}
where the summation over all Cartesian indices is implied. Thus for a 
given symmetry direction, $b$ will be a function of $b_{ijkl}$, which
are components of a fourth rank tensor ${\mathrm{\bf b}}$. 
 
The first term $F_0$ in Eq.~(\ref{f})
contains the contribution of all other degrees of freedom such as 
electrons, phonons, etc. The coefficient $a=a(T)$ is related to 
the inverse quasistatic field-cooled susceptibility $\chi_1$, namely,
$a=(\varepsilon_0 \chi_1)^{-1}$. The susceptibility $\chi_1$ can be 
calculated from the static spherical random bond--random field 
(SRBRF) model of relaxor ferroelectrics \cite{PB1,PBBG} and has the 
general form,
\begin{equation}
\chi_1=\frac{\Theta(1-q)}{T-T_0(1-q)},
\label{chi1} 
\end{equation}
which is well known from the theory of spin glasses and was found
empirically to hold in the case of relaxors \cite{V1}. Here, ${\Theta}$
is the Curie constant and $T_0$ a measure of the average interaction 
between the elementary dipolar entities in the system. In relaxors,
these are known to be the polar nanoregions (PNRs), which
are formed below the Burns temperature \cite{C1}. Finally, $q=q(T)$ 
is the dipolar glass order parameter, which is nonzero at 
all temperatures due to the presence of quenched random electric 
fields \cite{WKG}. In zero applied field, the order parameter $q$ 
is determined by the real solution 
of the following algebraic equation \cite{PB1,PBBG}:
\begin{equation}
q=(J/kT)^2(q+\Delta/J^2)(1-q)^2.
\label{q}
\end{equation}
The parameter $J$ is defined in terms of the variance $J^2/N$ 
of the infinitely ranged random interactions of a spin-glass type, 
and $\Delta$ the variance of local random fields. For PMN, the 
estimated values are $J/k\sim 217$~K and $\Delta/J^2=0.001$, 
whereas $T_0$ in Eq.~(\ref{chi1}) is of the order 
$kT_0\equiv J_0\sim 0.9 J$ \cite{BKPBL}.

The parameters $b,c,...$ in Eq.~({\ref{f}) are related to the 
nonlinear susceptibilities $\chi_3$, $\chi_5$,..., which are defined 
as usual by the expansion 
$P/\varepsilon_0=P_s + \chi_1 E+\chi_3 E^3 + \chi_5 P^5 +\cdots$.
In relaxors, the spontaneous polarization vanishes, thus by definition 
$P_s\equiv 0$, and the Landau coefficient $b$ is given by 
$b=-\chi_3/(\varepsilon_0^3\chi_1^4)$. It should be emphasized that
in general both $b$ and $\chi_3$, as well as higher order Landau 
coefficients in Eq.~({\ref{f}), depend on the direction of the 
field $\vec{E}$ due to the anisotropy term (\ref{f4}). 

The SRBRF model was originally introduced for an ideal isotropic relaxor 
system \cite{PB1} in a {\it rigid} environment. Thus the nonlinear 
susceptibility $\chi_3$ derived from it is independent of the 
orientation and $\chi^{rigid}_3<0$. Consequently, $b_{rigid}>0$, i.e.,  
$b_{rigid}$ is a positive scalar. Experiments on various relaxors have 
shown that $\chi_3$ can either be positive or negative, depending 
on the particular system studied and on the field 
orientation \cite{K2,TG}.

In order to derive a more general version of the model capable of
reproducing the observed anisotropy of the coefficient $b$, we 
introduce a coupling between the polarization $P$ and the strain 
tensor $u_{ij}$ (or the {\it internal} stress tensor $X_{ij}$). 
This suggests that we should consider the stress dependence 
of the Landau coefficients in Eq.~(\ref{f}). Focusing on the $P^2$-term, 
we first introduce a generalized Landau coefficient   
$a_{kl}=\varepsilon_0^{-1}(\chi_1^{-1})_{kl}$. Next, by 
expanding $a(X_{ij},T)_{kl}$ to linear order in $X_{ij}$, we replace
the $P^2$-term in $F$ by
\begin{equation}
\frac{1}{2\varepsilon_0}\left[\chi_1^{-1}\delta_{kl}
+\left(\frac{\partial(\chi_1^{-1})_{kl}}
{\partial X_{ij}}\right)_{E,T}X_{ij}+\cdots\right]P_kP_l.
\label{aX}
\end{equation}
The partial derivative is related to the electrostriction 
tensor $Q_{ijkl}$, namely, \cite{U}  
\begin{equation}
Q_{ijkl}=\frac{1}{2\epsilon_0}\left(\frac{\partial(\chi_1^{-1})_{kl}}
{\partial X_{ij}}\right)_{E,T}.
\label{Q}
\end{equation}
By adding the elastic energy, the free energy $F$ acquires an
additional term, which can be written as
\begin{equation}
F_X=\mathrm{\bf X}\cdot\mathrm{\bf Q}
\cdot\mathrm{\bf P}^2+\frac{1}{2}\mathrm{\bf X}
\cdot\mathrm{\bf C}^{-1}\cdot\mathrm{\bf X}.
\label{FX}
\end{equation}
Here, $\mathrm{\bf C}$ is the elastic constant tensor and
$(\mathrm{\bf P}^2)_{ij}=P_iP_j$.

Minimizing $F_X$ with respect to $X_{ij}$ 
at constant temperature and field, 
we formally recover the free energy (\ref{f}), however, the 
quartic term is now replaced by the general expression (\ref{f4}) with
\begin{equation}
b_{ijkl} = b_{rigid}\delta_{ij}\delta_{kl} + B_{ijkl},
\label{b}
\end{equation}
where the fourth rank tensor $\mathrm{\bf B}$ is given by 
\begin{equation}
\mathrm{\bf B} = -2\mathrm{\bf Q}\cdot\mathrm{\bf C}\cdot\mathrm{\bf Q}.
\label{B}
\end{equation}
In relaxors, the electrostriction effect is usually large and 
the magnitude of the tensor components $B_{ijkl}$ may exceed 
the value of $b_{rigid}$. Obviously, the sign of $B_{ijkl}$ will in 
general depend on the balance between the individual components
of $Q_{ijkl}$ and $C_{ijkl}$. Thus the resulting value of $b$ 
for a symmetry direction can either be positive or negative.

\begin{table*}
\caption{\label{tab:table1} 
Values of $C_{ij}$ and $Q_{ij}$ used to calculate $B_{ij}$ from
Eq.~(\ref{B}) and $B^{[p]}$ from Eq.~(\ref{bp}).}
~~~~~\\
\begin{ruledtabular}
\begin{tabular}{|cc|cc|cc|c|}
$C_{11}$ & 155.3$^{a,b}$ &  $C_{12}$ & 78.4$^{a}$ ~~~~~& $C_{44}$ 
& 68.3$^{a,b}$ 
~~~~~& [GPa]~~~~~~~ \\
\hline
$Q_{11}$ & 2.52$^c$ & $Q_{12}$ & -0.96$^c$ ~~~~~& $Q_{44}$ & 6.96$^d$ 
~~~~~& [10$^{-2}$m$^4$C$^2$]~~~~~~~\\
\hline                             
$B_{11}$ & 0.141 & $B_{12}$ & 0.483 ~~~~~& $B_{44}$ & -1.654  
~~~~~& [10$^8$Vm$^5$C$^3$]~~~~~~~\\
  \hline
$B^{[100]}$ & 0.141 & $B^{[110]}$ & -0.84 ~~~~~~~& $B^{[111]}$ & 
-1.836 ~~~~~& [10$^8$Vm$^5$C$^3$]~~~~~~~ \\
\end{tabular}
\end{ruledtabular}
$^a$Reference~\cite{LFGS}; $^b$Reference~\cite{TSS}; 
$^c$References~\cite{U,SN}; $^d$estimated~\cite{SN} 
\end{table*}

The generalized Landau free energy (\ref{f}) with $a(T)$ given by
the SRBRF model and the $P^4$-term having 
the form (\ref{f4}), and with quartic coefficients (\ref{b}) 
given by Eq.~(\ref{b}), will be referred to 
as the Compressible Spherical Dipolar Glass (CSDG) Model.

We can evaluate the coefficients $B_{ijkl}$ for the cases where values
of $Q_{ijkl}$ and $C_{ijkl}$ are explicitly known. In Table I, these
are listed for the PMN crystal using the Voigt notation, i.e.,
$B_{1111}=B_{11}$, etc. The value of $Q_{44}$ can be estimated from 
the relation \cite{SN}  $Q_{44}=(Q_{11}-Q_{12})/2$. From Eq.~(\ref{B})
we can then calculate the coefficients $B_{11}$, $B_{12}$, and 
$B_{44}$ (see Table I). For a symmetry direction $p$, where $p$ refers 
to [100], [110], or [111], the Landau coefficient $b=b^{[p]}$ 
in Eq.~(\ref{f}) can be expressed in terms of $b_{ij}$ as follows: 
\begin{subequations}
\label{bp} 
\begin{eqnarray} 
&b^{[100]}=b_{11};~~~~~~~~~~~~~~~~~~~~ \\ 
\label{bpa}
&b^{[110]}=\frac{1}{2}(b_{11}+b_{12}+2b_{44});\\
\label{bpb}
&b^{[111]}=\frac{1}{3}(b_{11}+2b_{12}+4b_{44}).
\label{bpc}
\end{eqnarray}
\end{subequations}

We can now write $b^{[p]}=b^{[p]}_{rigid}+B^{[p]}$, where $b_{rigid}>0$.  
Thus the rigid model always yields an isotropic, positive contribution 
to $b^{[p]}$, however, the new term $B^{[p]}$ is in general anisotropic. 
The corresponding values of $B^{[p]}$ for PMN are listed in Table~I. 
While $B^{[100]}>0$, we can see that $B^{[110]}$ and $B^{[111]}$ are both
negative. Thus, $b^{[100]}>0$ and $\chi_3^{[100]}<0$. On
the other hand, if $\vert B^{[p]}\vert > b_{rigid}$ for
$p =[110]$ and [111], the values of $b^{[110]}$ and $b^{[111]}$ 
will be negative, implying $\chi_3^{[110]}>0$ and $\chi_3^{[111]}>0$,
respectively. It is interesting to note that Tagantsev and
Glazounov \cite{TG} observed $\chi_3^{[111]}>0$, but $\chi_3^{[100]}<0$
in PMN, in agreement with the above conclusions. Quasistatic 
measurements of polarization versus field along [111] and [100]
directions also agree with these results \cite{K2}. As shown below, 
the sign of $b^{[p]}$ has important consequences for the existence 
of field-induced critical points for fields along the direction $[p]$. 

\section{Field induced critical points}

The temperature and field dependence of the dielectric polarization
during a field-cooled (or field-heated) quasi-stationary process is
calculated by minimizing numerically the free energy (\ref{f}). 
This procedure automally selects the correct solution of the
minimization condition $(\partial F/\partial P)_{E,T}=0$. The 
parameter $a=a(E,T)$ is calculated from the SRBRF model \cite{PKBZ}, 
while $b$ and $c$ are treated as free parameters. To ensure stability, 
we assume that $c>0$ and consider two cases, $b>0$ and $b<0$. For $b>0$,
$P(E,T)$ is found to increase monotonically with $E$ and decrease 
with $T$, and no critical singularities of the susceptibility
can be expected. For $b<0$, however, $P(E)$ makes a discontinuos jump 
at some value of $E$ at low temperatures. As $T$ increases, the 
jump becomes smaller and finally disappears at the critical 
point $E_{CP},T_{CP}$, where the slope of $P(E)$ is infinite.
This is illustrated in Fig.~2 for $b=-0.2$ and $c=0.08$, corresponding 
to PMN [110]. A similar behavior of $P(T)$ had been obtained earlier 
for PMN [111], where $b=-1/3$ and $c=|b|$ \cite{PKBZ}. 


\begin{figure}
\resizebox{20.4pc}{!}{\includegraphics{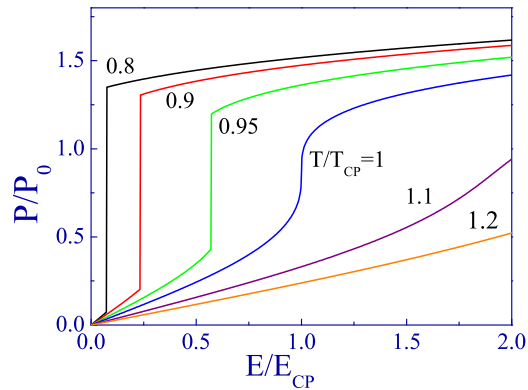}}
\caption{Field dependence of $P(E)$ for a relaxor with $b<0$ 
and several values of temperature $T$ close to the
critical temperature $T_{CP}$, obtained by minimizing the free 
energy (\ref{f}). Note that these calculations are only valid in
the ergodic region above the freezing line shown in Fig.~1.}
\label{fig2}
\end{figure} 


The coordinates of the critical point 
are determined from the relations \cite{K6,PKBZ}
\begin{equation}
a(T_{CP})=\frac{9b^2}{20c}; \;\; 
E_{CP}= \frac{6b^2}{25c}P_{CP},
\label{CP}
\end{equation}
where $P_{CP}=\sqrt{-3b/(10c)}$ is the polarization at the
critical point. The critical exponents at the 
field-induced critical point differ from the usual mean field 
exponents for ferroelectrics in zero field \cite{PKBZ}.

The $E$-$T$ phase diagram for PMN corresponding to cooling in 
a field along the [110] direction is plotted in Fig.~1 using the data 
points from Ref.~\cite{NWPK}. Similar phase diagrams were obtained
earlier for PMN [100] and [111] \cite{K2,C2}. For $E\| [100]$ no
critical point was found, but the phase diagram for the [111] direction
was shown to be analogous to the [110] case, in agreement with the 
above estimates for $b^{[p]}$. A similar conclusion had been reached 
earlier by Zhao et al. \cite{z1}.


\begin{figure}
\resizebox{30.4pc}{!}{\includegraphics{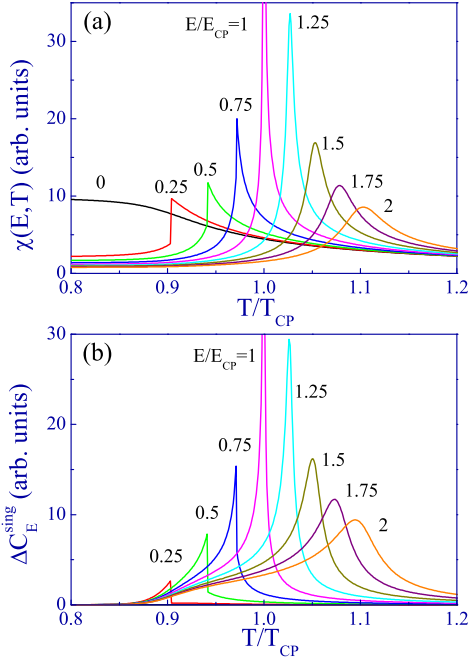}}
\caption{(a) Calculated temperature dependence of the suceptibility
$\chi(E,T)$ for a set of field values $E/E_{CP}$, as indicated.
(b) Same, but for the singular part of the specific heat 
from Eq.~(\ref{Cs}).
}
\label{fig3}
\end{figure}
 

The existence of field-induced phase transitions and critical points
has recently been confirmed in PMN [110] by measuring 
the specific heat using high-resolution calorimetry \cite{NWPK}. 
The excess specific heat $\Delta C_E(T)$, which is due to the 
contribution of the dipolar degrees of freedom, namely, PNRs can be
derived from the free energy Eq.~(\ref{f}) by applying the thermodynamic 
relation for the entropy $S=-(\partial F/\partial T)_E$. We can 
write $S=S_0+S_{dip}$, where $S_0=-(\partial F_0/\partial T)_E$
and the dipolar part $S_{dip}$ is defined as the contribution
of all $P$-dependent terms in Eq.~(\ref{f}),  
\begin{equation}
S_{dip}=-\left(\frac{1}{2}a_1P^2+\frac{1}{4}b_1P^4+\frac{1}{6}c_1P^6
+\cdots\right),
\label{Sdip}
\end{equation}
where $a_1\equiv da/dT$, $b_1\equiv db/dT$, etc., and the condition
$(\partial F/\partial P)_E=0$ has been applied \cite{PKBZ}. The 
dipolar excess specific heat capacity at constant field is given by 
$\Delta C_E=T(\partial S_{dip}/\partial T)_E$, and at constant polarization
similarly by $\Delta C_P=T(\partial S_{dip}/\partial T)_P$. These 
two quantities are related by the standard thermodynamic relation   
\begin{equation}
\Delta C_E=\Delta C_P+T\chi(E,T)[(\partial E/\partial T)_P]^2,
\label{CEP}
\end{equation}
where $\chi(E,T)=(\partial P(E,T)/\partial E)_T$ is the field-dependent 
susceptibility. The partial derivative in Eq.~(\ref{CEP}) can be 
evaluated from the equation of state $E=aP+bP^3+cP^5+\cdots$, i.e., 
\begin{equation}
(\partial E/\partial T)_P=a_1P+b_1P^3+c_1P^5+\cdots.
\label{ET}
\end{equation}
To calculate the expression on the right hand side we would, therefore, 
need to know the temperature dependence of the coefficients $a,b,c,$ etc.
In practice, $a(T)$ is known from the SRBRF model through the relation 
$a=(\varepsilon_0 \chi_1)^{-1}$, and $a_1(T)$ follows from the temperature
derivative of $\chi_1(T)$. On the other hand, $b(T)$ and $c(T)$ could,
in principle, be estimated from the nonlinear susceptibilities as given
by the SRBRF model and the $\mathrm{\bf Q}$ and $\mathrm{\bf C}$ tensors. 
In the following we will simply assume that $b$ and $c$ are effectively 
constant in the temperature range of interest, and 
thus $b_1$ and $c_1$ will be neglected. 


\begin{figure}
\resizebox{30.4pc}{!}{\includegraphics{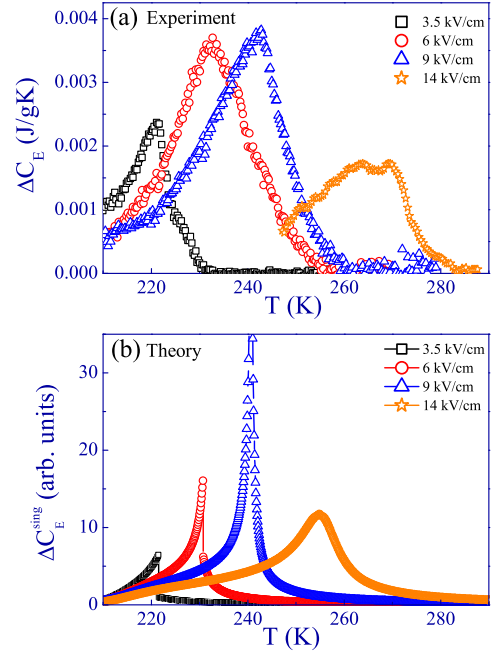}}
\caption{
(a) Experimental data showing the specific heat anomalies occurring
in PMN [110] for four selected values of the electric field,
obtained by high-resolution calorimetry \cite{NWPK}.
(b) Calculated temperature dependence of the singular
part of the excess specific heat $C^{sing}_E$ for a relaxor 
with $b=-0.2$ and $c=0.08$ and the same field values as used
in the experiment. The remaining parameters are taken from 
Ref.~\cite{PKBZ}.
}
\label{fig4}
\end{figure}
 
The calculated temperature dependence of $\chi(E,T)$ is shown in Fig. 3a
for a set of values $0\le E\le 2E_{CP}$. Here we used the parameter values 
of $J$, $J_0$, and $\Delta$ determined earlier from the dielectric 
data \cite{BKPBL}, and the remaining parameters were chosen as 
$b=-0.2$ and $c=0.08$. At $E=0$, the zero-field cooled
susceptibility $\chi_1$ is recovered. For $0<E\le E_{CP}$, 
$\chi(E,T)$ exhibits a jump at the first order transitions and 
diverges at the critical point. For $E>E_{CP}$, however, $\chi(E,T)$ 
is characterized by rounded maxima, in accordance with the smooth 
behavior of $P(E,T)$ in the supercritical regime.

The first term in Eq.~(\ref{CEP}) is readily shown to be   
\begin{equation}
\Delta C_P=-T\left(\frac{1}{2}a_2P
+\frac{1}{4}b_2P^3+\frac{1}{4}c_2P^5+\cdots\right),
\label{CPP}
\end{equation}
where $a_2\equiv da_1/dT$ etc. Again, the corresponding derivatives of  
$b,c,...$ are not known, and we will neglect them. The quantity
$a_2$ can, however, be calculated from $a(T)$. It shows a 
sharp peak around the static "freezing" temperature 
$T_f=(J^2+\Delta)^{1/2}/k$, but is rather small elsewhere. We may  
conclude that $\Delta C_P$ will be nonsingular for all values of
$E,T$, however, its precise behavior could only be determined if
the values of $b(T)$ and $c(T)$ were known. 

The temperature dependence of the singular part of $\Delta C_E(T)$
can be calculated from Eqs.~(\ref{CEP}) and (\ref{ET}), namely,
\begin{equation}
\Delta C^{sing}_E\cong T\chi(E,T)(a_1P)^2,
\label{Cs}
\end{equation}
and is displayed in Fig.~3b for the same set of parameters as 
in Fig.~3a. The positions of the singularities of $\Delta C^{sing}_E(T)$ 
coincide with those of $\chi(E,T)$, however, the direction of the
jumps is reversed due to the last factor in Eq.~(\ref{Cs}).

The experimental data for $\Delta C_E(T)$ in PMN [110] obtained by
high-resolution calorimetry \cite{NWPK} are shown in Fig.~4a for four 
discrete values of the electric field $E$.
For comparison, the theoretical prediction for the singular part of
$\Delta C_E(T)$, calculated from Eq.~(\ref{Cs}) at the same field 
values, is plotted in Fig.~4b. The experimental values for the 
critical field and the critical temperature are given by 
$E_{CP}\cong 8$~kV/cm and $T_{CP}\cong 240$~K, respectively \cite{NWPK}. 

The predicted behavior of $\Delta C^{sing}_E(T)$ qualitatively agrees 
with the experimental values of $\Delta C_E(T)$, however, the experimental 
anomalies appear to be broader than the calculated ones. There are 
several reasons for this broadening, for example, finite size effects, 
structural inhomogeneities, and slow relaxation. 

It should be stressed that at zero field ($E=0$), no anomalies in
$\Delta C_E(T)$ were found in the entire temperature range studied,
in accordance with the CSDG model. This contrasts with the so-called
random field scenario, according to which PMN [110] is assumed to 
undergo a ferroelectric phase transition at $E=0$. This point has 
been discussed in more detail in Ref.~\cite{NWPK}.

\section{Conclusions}

We have shown that the electrostrictive coupling between 
the dielectric polarization $P$ and the strain or stress tensor 
fluctuations in a relaxor ferroelectric gives rise to an 
anisotropy of the $P^4$-term in the Landau free energy. 
For a given symmetry direction of the applied 
field the effective Landau coefficient $b$ may become negative, thus
leading to the field-induced critical points. The compressible spherical
dipolar glass model predicts singularities of the dipolar specific 
heat near critical points, in agreement with high-resolution 
calorimetry experiments in PMN [110] \cite{NWPK}.

{\bf Acknowledgment.}
This work was supported by the Slovenian Research Agency through
Grants P1-0044, P1-0125, J1-0155, and J1-2015. 




\newpage



\newpage


\end{document}